\documentclass [aps, showpacs, superscriptaddress] {revtex4}
\usepackage{graphics}
\newcommand{\sinp}
{\affiliation{Theoretical Condensed Matter Physics Division
 and Centre for Applied Mathematics and Computational Science,\\
 Saha Institute of Nuclear Physics,
 Sector - 1, Block - AF, Bidhannagar, Kolkata 700 064, India}}
\newcommand{\gc}
{\affiliation{Physics Department, Gurudas College, Narkeldanga,
 Kolkata 700 054, India}}

\begin{document}

\title{A common mode of origin of power laws in models of market and
earthquake}

\author{Pratip Bhattacharyya}
\email{pratip.bhattacharyya@saha.ac.in}
\sinp
\gc

\author{Arnab Chatterjee}
\email{arnab.chatterjee@saha.ac.in}
\sinp

\author{Bikas K. Chakrabarti}
\email{bikask.chakrabarti@saha.ac.in}
\sinp

\date{November 04, 2005}

\begin{abstract}
We show that there is a common mode of origin for the power laws observed
in two different models: (i) the Pareto law for the distribution of money
among the agents with random saving propensities in an ideal gas-like
market model and (ii) the Gutenberg-Richter law for the distribution
of overlaps in a fractal-overlap model for earthquakes. We find that
the power laws appear as the asymptotic forms of ever-widening log-normal
distributions for the agents' money and the overlap magnitude respectively.
The identification of the generic origin of the power laws helps in better
understanding and in developing generalized views of phenomena
in such diverse areas as economics and geophysics.
\end{abstract}


\maketitle

\section{Introduction}

\indent Recently we have shown that the Pareto law \cite{Pareto:1897}
appears asymptotically ($m \to \infty$) in the distribution of money $m$
among the agents in the steady state of a trading market model:
\begin{equation}
P(m) \sim m^{-(1+\nu)}, \ \nu = 1
\label{eq:Pareto-law}
\end{equation}

\noindent when the agents have random saving propensities
\cite{EWD:2005, Chatterjee:2004}. The market is modeled as an ideal
gas where each molecule is identified with an agent, with the
additional attribute that each agent has a random saving propensity,
and each trading event between two agents is considered to be an elastic
or money conserving collision between two molecules. In another model ---
a geometric model for earthquakes \cite{Chakrabarti:1999} --- we have
shown that a power law similar to the Gutenberg-Richter law
\cite{Gutenberg:1944, Gutenberg:1954} appears in the asymptotic distribution
of the overlap $S$ between two dynamically intersecting Cantor sets:
\begin{equation}
G(S) \sim S^{-\gamma}, \ \gamma = 1.
\label{eq:Gutenberg-Richter-law}
\end{equation}

\noindent Since a geological fault is formed of a pair of fractal rock
surfaces that are in contact and in relative motion, it is modeled by
a pair of overlapping Cantor sets (the simplest known fractal), one
shifting over the other. The overlap between the two Cantor sets
represents the area of contact between the two surfaces of the fault
and hence it is proportional to the energy released in an earthquake
resulting from ruptures in the regions of contact.

\indent In both the models we get simple power laws with the exponents
$\nu = \gamma = 1$. Although these have been obtained separately for
the two models, using both numerical and analytic methods, we show
here that the two cases have a common feature that results in a common
mode of origin of the power laws observed in the distribution of
money $m$ and fractal overlap $S$. The derivation of the power laws
presented here shows that the common feature is a log-normal
distribution in which the normal factor spreads indefinitely thus leaving
the power-law factor to dominate the asymptotic distribution.

\section{The ideal gas market model}

\indent Let us first consider the ideal gas model of an isolated economic
system --- that we refer to as the \lq market\rq~ --- in which the total
money $M$ and the total number of agents $N$ are both constant; there is
neither any production nor any destruction of money within the market and
no migration of agents occurs between the market and its environment.
The only economic activity allowed in the market is trading among the
agents. Each agent $i$ possesses an amount of money $m_i(t)$ at time $t$.
The time $t$ is discrete and each event of trading is counted as a unit
time step. In an event of trading (shown schematically in Fig. 1) a pair
of agents $i$ and $j$ randomly redistribute their money between themselves
such that their total money is conserved and none of the two agents emerges
from the trading process with negative money (i.e., debt is not allowed):

\begin{eqnarray}
m_i(t+1) + m_j(t+1) = m_i(t) + m_j(t), & & \\
m_i(t) \geq 0 \ \mathrm{for} \ \mathrm{all} \ i \
 \mathrm{at} \ \mathrm{all} \ t.
 \nonumber
\label{eq:local-conservation}
\end{eqnarray}

\begin{figure}
\centering{
\resizebox*{7.0cm}{!}{\includegraphics{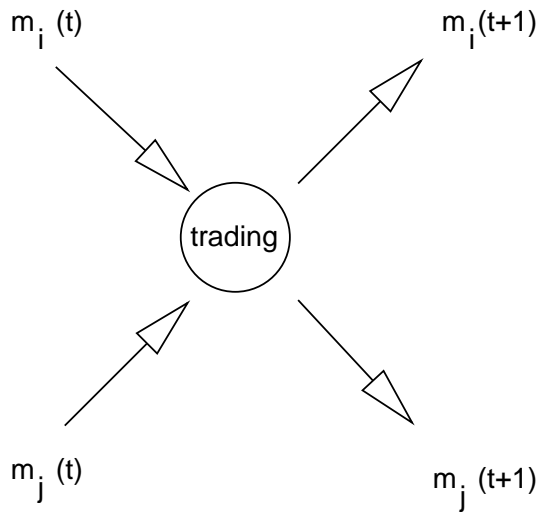}}
}
\caption{Schematic diagram of the trading process. Agents $i$ and $j$
redistribute their money in the market: $m_i(t)$ and $m_j(t)$, their
respective money before trading, changes over to $m_i(t+1)$ and $m_j(t+1)$
after trading.}
\label{fig1:trading-scheme}
\end{figure}

\indent It has already been shown that in the steady state market
($t \to \infty$) the money $m$ with the individual agents follow the
Gibbs distribution \cite{Chak1995-Yak:2000}:
\begin{equation}
P(m) = {1 \over T} \exp{\left ( - {m \over T} \right )}; \ T = {M \over N}
\label{eq:Gibbs}
\end{equation}

\noindent when there is no restriction on the amount of money each agent
can trade with except that it must satisfy the conditions of Eq.
(\ref{eq:local-conservation}). Here $T$ represents the economic equivalent
of temperature and is defined as the average money per agent in the market.
If each agent saves a fraction $\lambda$ ($0 \leq \lambda < 1$) of its
own money at every trading and $\lambda$ is the same for all agents at
all time steps, the individual money with the agents in the steady state
follows the Gamma distribution \cite{Chak2000-Das2003-Pat:2004}.
If we consider the effect of randomly distributed saving fraction
$\lambda_i$ among the agents $i$, the money distribution in the steady
state assumes the form of the Pareto law. The evolution of the agents'
money in a trading can be written as
\begin{equation}
m_i(t+1) = \lambda_i m_i(t) + \epsilon_t \left [ (1 - \lambda_i) m_i(t)
                                  + (1 - \lambda_j) m_j(t)  \right ]
\label{eq:evolution-i}
\end{equation}
\noindent and
\begin{equation}
m_j(t+1) = \lambda_j m_j(t) + (1-\epsilon_t)
               \left [ (1 - \lambda_i) m_i(t)
                       + (1 - \lambda_j) m_j(t)  \right ]
\label{eq:evolution-j}
\end{equation}

\noindent where $\lambda_i$ and $\lambda_j$ are the saving fractions of
agents $i$ and $j$ respectively. The saving fractions $\lambda_i$ are
quenched, i.e., fixed in time for each agent $i$ and are distributed
randomly and uniformly (like white noise) on the interval $[0,1)$.
The random division of the total traded money is given by the number
$0 \leq \epsilon_t \leq 1$ that varies randomly with the trading events
$t$. The money distribution in the steady state is found to have a long
power-law tail (shown in Fig.2) that fits with the Pareto law for
$\nu = 1$ \cite{Chatterjee:2004}. We also have analytic proofs
\cite{EWD:2005,Chatterjee:2005, Repetowicz:2005} of the
Pareto distribution of money observed in this random-saving gas-like
model; all these proofs proceed by formulating the trading events as
scattering processes and show that the Pareto distribution is a steady
state solution of the scattering problem.

\begin{figure}
\centering{
\resizebox*{10.0cm}{!}{\includegraphics{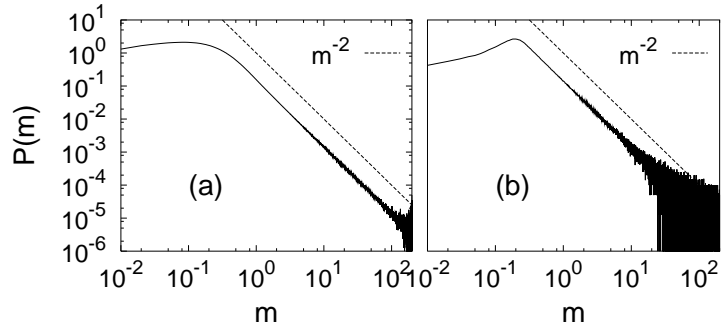}}
}
\caption{The money distribution $P(m)$ for (a) random $\epsilon_t$
and (b) $\epsilon_t = 1/2$. The power law in both cases has the same
exponent: $P(m) \sim m^{-2}$. Numerical simulation has been done for
a system of $N=200$ agents.}
\label{fig2:market-power-law}
\end{figure}

\indent Here we give a simple derivation of the asymptotic distribution
of money in the steady state of the market model using an argument of the
mean-field type, thereby avoiding the intricacies of the previous proofs.
In our approach the money redistribution equations (\ref{eq:evolution-i})
and (\ref{eq:evolution-j}) are reduced to a single stochastic map by
taking the product of the two equations:
\begin{equation}
m_i(t+1) m_j(t+1) = \alpha_i (\epsilon_t,\lambda_i) m_i^2(t)
                    + \alpha_j (\epsilon_t,\lambda_j) m_j^2(t)
                + \alpha_{ij} (\epsilon_t,\lambda_i, \lambda_j) m_i(t) m_j(t).
\label{eq:evolution-ij}
\end{equation}

\noindent Now we introduce a mean-field-like approximation by replacing
each of the quadratic quantities $m_i^2$, $m_j^2$ and $m_i m_j$ by a mean
quantity $m^2$. Therefore Eq. (\ref{eq:evolution-ij}) is replaced by its
mean-field-like approximation
\begin{equation}
m^2(t+1) = \eta(t) m^2(t)
\label{eq:stochastic-map}
\end{equation}

\noindent where $\eta(t)$ is an algebraic function of $\lambda_i$,
$\lambda_j$ and $\epsilon_t$; it has been observed in numerical
simulations of the model that the value of $\epsilon_t$, whether it is
random or constant, has no effect on the steady state distribution
\cite{Chatterjee:2005} (illustrated in Fig. 2) and the time dependence
of $\eta(t)$ results from the different values of $\lambda_i$ and
$\lambda_j$ encountered during the evolution of the market. Denoting
$\log (m^2)$ by $x$, Eq. (\ref{eq:stochastic-map}) can be written as:
\begin{equation}
x(t+1) = x(t) + \delta(t),
\label{eq:random-walk}
\end{equation}

\noindent where $\delta(t) = \log \eta(t)$ is a random number that changes with
each time-step. The transformed map (Eq. \ref{eq:random-walk}) depicts a
random walk and therefore the `displacements' $x$ in the time interval
$[0,t]$ follows the normal distribution
\begin{equation}
\mathcal{P}(x) \sim \exp \left ( -{x^2 \over t} \right ).
\label{eq:RW-distribution}
\end{equation}

\noindent Now
\begin{equation}
\mathcal{P} (x) \mathrm{d} x \equiv P(m) \mathrm{d} m^2
\label{eq:equivalence1}
\end{equation}

\noindent where $P(m)$ is the log-normal distribution of $m^2$:
\begin{equation}
P(m) \sim {1 \over m^2} \exp \left [ -{\left ( \log(m^2) \right )^2 \over t}
 \right ].
\label{eq:log-normal1}
\end{equation}

\noindent The normal distribution in Eq. (\ref{eq:RW-distribution}) spreads
with time (since its width is proportional to $\sqrt{t}$) and so does the
normal factor in Eq. (\ref{eq:log-normal1}) which eventually becomes a very
weak function of $m$ and may be assumed to be a constant as $t \to \infty$.
Consequently $P(m)$ assumes the form of a simple power law:
\begin{equation}
P(m) \sim {1 \over m^2} \ \mathrm{for} \ t \to \infty,
\label{eq:power-law1}
\end{equation}

\noindent that is clearly the Pareto law (\ref{eq:Pareto-law}) for the model.

\section{The fractal-overlap model of earthquake}

\indent Next we consider a geometric model \cite{Chakrabarti:1999} of the
fault dynamics occurring in overlapping tectonic plates that form the earth's
lithosphere. A geological fault is created by a fracture in the earth's
rock layers followed by a displacement of one part relative to the other.
The two surfaces of the fault are known to be self-similar fractals. In
this model a fault is represented by a pair of overlapping identical
fractals and the fault dynamics arising out of the relative motion of the
associated tectonic plates is represented by sliding one of the fractals
over the other; the overlap $S$ between the two fractals represents the
energy released in an earthquake whereas $\log S$ represents the magnitude
of the earthquake. In the simplest form of the model each of the two
identical fractals is represented by a regular Cantor set of fractal
dimension $\log 2 / \log 3$. This is the only exactly solvable model for
earthquakes known so far. The exact analysis of this model
\cite{Bhattacharyya:2005} for a finite generation $n$ of the Cantor sets
with periodic boundary conditions showed that the probability of
the overlap $S$, which assumes the values $S=2^{n-k} (k=0, \ldots , n)$,
follows the binomial distribution $F$ of $\log_2 S = n-k$:

\begin{eqnarray}
\lefteqn{\Pr \left ( S=2^{n-k} \right )
 \equiv \Pr \left ( \log_2 S = n-k  \right )} \nonumber \\
 & &= \left ( \begin{array}{c} n\\ n-k \end{array} \right )
      \left ( {1 \over 3}  \right )^{n-k} \left ( {2 \over 3}  \right )^k
      \equiv F(n-k).
\label{eq:binomial-regular}
\end{eqnarray}

\begin{figure}
\centering{
\resizebox*{7.0cm}{!}{\includegraphics{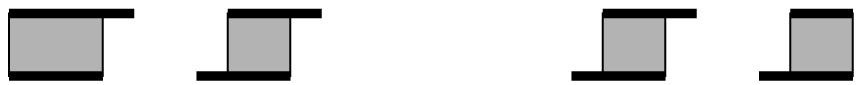}}
\vskip 0.5cm
\resizebox*{7.0cm}{!}{\includegraphics{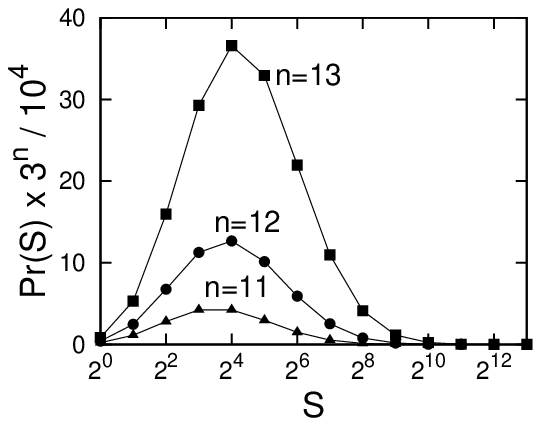}}
}
\caption{The second generations ($n=2$) of two Cantor sets with periodic
boundary conditions, one shifting uniformly over the other. The total
length of the shaded regions gives the overlap $S$ between the two sets
at any instant of time. The distribution of $\log_2 S$ is given by Eq.
(\ref{eq:binomial-regular}) that is seen to approach a log-normal
distribution of $S$ (Eq. \ref{eq:log-normal2}) with increasing $n$
and will eventually assume the asymptotic power law form
(Eq. \ref{eq:power-law2}) as $n \to \infty$.}
\label{fig3:cantor}
\end{figure}

\noindent Since the index of the central term (i.e., the term for the most
probable event) of the above distribution is $n/3 + \delta$,
$-2/3 < \delta < 1/3$, for large values of $n$ Eq. (\ref{eq:binomial-regular})
may be written as
\begin{equation}
F \left ( {n \over 3} \pm r \right ) \approx \left ( \begin{array}{c}
                                            n\\ n \pm r \end{array} \right )
                            \left ( {1 \over 3}  \right )^{{n \over 3} \pm r}
                            \left ( {2 \over 3}  \right )^{{2n \over 3} \mp r}
\label{eq:binomial-central}
\end{equation}

\noindent by replacing $n-k$ with $n/3 \pm r$. For $r \ll n$,
we can write the normal approximation to the above binomial distribution as
\begin{equation}
F \left ( {n \over 3} \pm r  \right ) \sim {3 \over \sqrt{2 \pi n}}
                             \exp{ \left ( -{9r^2 \over 2n} \right )}
\label{eq:normal-approx}
\end{equation}

\noindent Since $\log_2 S = n-k = {n \over 3} \pm r$, we have
\begin{equation}
F \left ( \log_2 S \right ) \sim {1 \over \sqrt{n}}
          \exp{\left [ - {\left ( \log_2 S \right )^2 \over n} \right ]},
\label{eq:normal-approx'}
\end{equation}

\noindent not mentioning the factors that do not depend on $S$. Now
\begin{equation}
F \left ( \log_2 S \right ) \mathrm{d} \left ( \log_2 S \right )
\equiv G(S) \mathrm{d} S
\label{eq:equivalence}
\end{equation}

\noindent where
\begin{equation}
G(S) \sim {1 \over S}
 \exp \left [ - {\left ( \log_2 S \right )^2 \over n} \right ]
\label{eq:log-normal2}
\end{equation}

\noindent is the log-normal distribution of $S$. As the generation index
$n \to \infty$, the normal factor spreads indefinitely (since its width is
proportional to $\sqrt{n}$) and becomes a very weak function of $S$ so that
it may be considered to be almost constant; thus $G(S)$ asymptotically
assumes the form of a simple power law with an exponent that is
independent of the fractal dimension of the overlapping Cantor sets:
\begin{equation}
G(S) \sim {1 \over S} \ \mathrm{for} \ n \to \infty.
\label{eq:power-law2}
\end{equation}

\noindent This is the Gutenberg-Richter law (Eq.
\ref{eq:Gutenberg-Richter-law}) for the fractal-overlap model of
earthquakes. It was also observed in numerical simulations
\cite{Pradhan:2003-04} that $G(S) \sim S^{-\gamma}, \ \gamma \approx 1$
for several other regular and random fractals, thus suggesting that the
exponent may be universal.

\indent The exact result of Eq. (\ref{eq:binomial-regular}) in Ref.
\cite{Bhattacharyya:2005} disagreed with the asymptotic power law of
Eq. (\ref{eq:power-law2}) obtained previously by renormalization group analysis
of the model for $n \to \infty$ in Ref. \cite{Chakrabarti:1999}. The
disparity between the two results had appeared because it was overlooked
that the former is the exact distribution of $\log_2 S$ whereas the latter
was the asymptotic distribution of $S$. However the above analysis shows
that the power law in Eq. (\ref{eq:power-law2}) is indeed the asymptotic
form of the exact result. This is qualitatively similar to what
is observed in the distribution of real earthquakes: the
Gutenberg-Richter power law is found to describe the distribution of
earthquakes of small and intermediate energies; however deviations from it
are observed for the very small and the very large earthquakes.

\indent The fact that the fractal-overlap model produces an asymptotic
power law distribution of overlaps suggests that the Gutenberg-Richter law
owes its origin significantly to the fractal geometry of the faults.
Furthermore, since this model contains the geometrical rudiments (i.e.,
the fractal overlap structure) of geological faults and it produces
an asymptotic distribution of overlaps that has qualitative similarity
with the Gutenberg-Richter law, we are inclined to believe that the entire
distribution of real earthquake energies is log-normal that is wide
enough for the Gutenberg-Richter power law to be observed over a large
range of energy values.

\section{Concluding remarks}
In the trading market model, we have shown that the money redistribution
equations for the individual agents participating in a trading process
can be reduced to a stochastic map in $m^2$ (Eq. \ref{eq:stochastic-map}).
Using the transformation $x = \log (m^2)$, the map was reduced to a
random walk in the variable $x$ and hence the distributions of $x$ and
$m^2 $ were found to be normal and log-normal respectively; in the steady
state, i.e., for $t \to \infty$, the latter was found to assume the form
of a power law identical to the Pareto law with the exponent $\nu = 1$.
Likewise, in the fractal-overlap model for earthquakes the distribution
of overlaps was found to be log-normal for large generation indices $n$
of the Cantor set and it further reduced asymptotically (as $n \to \infty$)
to a power law similar to the Gutenberg-Richter law for earthquake energies.
In both the cases, the original distribution of the relevant variable
($m^2$ and $S$) was log-normal in which the normal factor became a very
weak function of the variable in the asymptotically ($t \to \infty$
and $n \to \infty$ respectively), thus rendering a power-law form to
the distribution. Our derivations of the two power laws in the two
vastly different models also indicate the universality of the exponents
$\nu=\gamma=1$. In particular, the value of the Gutenberg-Richter exponent
$\gamma=1$ in the fractal-overlap model is clearly independent of the
dimension of the fractals used and therefore the result is of a general
nature. In the context of this paper it may be mentioned that in a
similar fashion Pietronero et al \cite{Pietronero:2001} found a common
mode of origin for the laws of Benford and Zipf.

\begin{acknowledgments}
BKC is grateful to S. Sinha for useful discussions. AC thanks E. Tosatti
and L. Pietronero for pointing out Ref. \cite{Pietronero:2001}.
\end{acknowledgments}


\begin{thebibliography}{19}

\bibitem{Pareto:1897}
V. Pareto, \textit{Cours d'economie Politique}, (F. Rouge, Lausanne, 1897).

\bibitem{EWD:2005}
\textit{Econophysics of Wealth Distributions}, Eds. A. Chatterjee, S.
Yarlagadda, and B. K. Chakrabarti (Springer-Verlag Italia, Milan, 2005).

\bibitem{Chatterjee:2004}
A. Chatterjee, B. K. Chakrabarti and S. S. Manna, Physica A
 \textbf{335} (2004) 155.

\bibitem{Chakrabarti:1999}
B. K. Chakrabarti and R. B. Stinchcombe, Physica A \textbf{270} (1999) 27.

\bibitem{Gutenberg:1944}
B. Gutenberg and C. F. Richter, Bull. Seismol. Soc. Am. \textbf{34} (1944)
 185.

\bibitem{Gutenberg:1954}
B. Gutenberg and C. F. Richter, \textit{Seismicity of the Earth},
 Princeton University Press, Princeton, 1954.

\bibitem{Chak1995-Yak:2000}
B. K. Chakrabarti and S. Marjit, Ind. J. Phys. B \textbf{69} (1995) 681;
A. A. Dr\u{a}gulescu and V. M. Yakovenko, Eur. Phys. J. B \textbf{17}
 (2000) 723.

\bibitem{Chak2000-Das2003-Pat:2004}
A. Chakraborti and B. K. Chakrabarti, Eur. Phys. J. B \textbf{17} (2000) 167;
A. Das and S. Yarlagadda, Phys. Scr. T \textbf{106} (2003) 39;
M. Patriarca, A. Chakraborti and K. Kaski, Phys. Rev. E \textbf{70}
 (2004) 016104.

\bibitem{Chatterjee:2005}
A. Chatterjee, B. K. Chakrabarti and R. B. Stinchcombe,
Phys. Rev. E \textbf{72} (2005) 026126.

\bibitem{Repetowicz:2005}
P. Repetowicz, S. Hutzler and P. Richmond, Physica A \textbf{356} (2005) 641.

\bibitem{Bhattacharyya:2005}
P. Bhattacharyya, Physica A \textbf{348} (2005) 199.

\bibitem{Pradhan:2003-04}
S. Pradhan, B. K. Chakrabarti, P. Ray and M. K. Dey, Phys. Scr.
 \textbf{T106} (2003) 77;
 S. Pradhan, P. Chaudhuri and B. K. Chakrabarti, in
 \textit{Continuum Models and Discrete Systems}, Ed. D. J. Bergman, E. Inan,
 Nato Sc. Series, Kluwer Academic Publishers (Dordrecht, 2004) pp. 245-250;
 arXiv:cond-mat/0307735.

\bibitem{Pietronero:2001}
L. Pietronero, E. Tosatti, V. Tosatti and A. Vespignani, Physica A
 \textbf{293} (2001) 297.

\end{thebibliography}
\end{document}